\documentclass[12pt]{article}
\usepackage{graphicx}
\usepackage{amsfonts}
\usepackage{amssymb}
\begin{document}
\def\journal{\topmargin .3in    \oddsidemargin .5in
        \headheight 0pt \headsep 0pt
        \textwidth 5.625in % 1.2 preprint size  %6.5in
        \textheight 8.25in % 1.2 preprint size 9in
        \marginparwidth 1.5in
        \parindent 2em
        \parskip .5ex plus .1ex         \jot = 1.5ex}
%       The hybrid style is adapted to print well on both
%       US paper and A4 paper by picking the correct dimensions.

\def\hybrid{\topmargin -20pt    \oddsidemargin 0pt
        \headheight 0pt \headsep 0pt
        \textwidth 6.25in
        \textheight 9.00in       % BS paper
        \marginparwidth .875in
        \parskip 5pt plus 1pt   \jot = 1.5ex}

%       The default is set to be hybrid

% \hybrid
\journal

\catcode`\@=11

\def\marginnote#1{}
%%%%%%%%%%%%%%%%%%%%%%%%%%%%%%%%%%%%%%%%%%%%%%%
%       The time macros where written by Jon Yamron
%
\newcount\hour
\newcount\minute
\newtoks\amorpm
\hour=\time\divide\hour by60
\minute=\time{\multiply\hour by60 \global\advance\minute by-\hour}
\edef\standardtime{{\ifnum\hour<12 \global\amorpm={am}%
        \else\global\amorpm={pm}\advance\hour by-12 \fi
        \ifnum\hour=0 \hour=12 \fi
        \number\hour:\ifnum\minute<10 0\fi\number\minute\the\amorpm}}
\edef\militarytime{\number\hour:\ifnum\minute<10 0\fi\number\minute}
%% FOLLOWING LINE CANNOT BE BROKEN BEFORE 70 CHAR
%% FOLLOWING LINE CANNOT BE BROKEN BEFORE 70 CHAR
%%%%%%%%%%%%%%%%%%%%%%%%%%%%%%%%%%%%%%%%%%%%%%%

\def\draftlabel#1{{\@bsphack\if@filesw {\let\thepage\relax
   \xdef\@gtempa{\write\@auxout{\string
      \newlabel{#1}{{\@currentlabel}{\thepage}}}}}\@gtempa
   \if@nobreak \ifvmode\nobreak\fi\fi\fi\@esphack}
        \gdef\@eqnlabel{#1}}
\def\@eqnlabel{}
\def\@vacuum{}
\def\draftmarginnote#1{\marginpar{\raggedright\scriptsize\tt#1}}

\def\draft{\oddsidemargin -.5truein
        \def\@oddfoot{\sl preliminary draft \hfil
        \rm\thepage\hfil\sl\today\quad\militarytime}
        \let\@evenfoot\@oddfoot \overfullrule 3pt
        \let\label=\draftlabel
        \let\marginnote=\draftmarginnote
   \def\@eqnnum{(\theequation)\rlap{\kern\marginparsep\tt\@eqnlabel}%
\global\let\@eqnlabel\@vacuum}  }

%       This defines the preprint style which is to be imprinted in
%       landscape mode. The command \preprint precedes the begin
%       document command.

\def\preprint{\twocolumn\sloppy\flushbottom\parindent 2em
        \leftmargini 2em\leftmarginv .5em\leftmarginvi .5em
        \oddsidemargin -.5in    \evensidemargin -.5in
        \columnsep .4in \footheight 0pt
        \textwidth 10.in        \topmargin  -.4in
        \headheight 12pt \topskip .4in
        \textheight 6.9in \footskip 0pt
        \def\@oddhead{\thepage\hfil\addtocounter{page}{1}\thepage}
        \let\@evenhead\@oddhead \def\@oddfoot{} \def\@evenfoot{} }

%       This sets the default for World Scientific proceedings or
%       metric size proceedings contributions.
\def\proceedings{\pagestyle{empty}
        \oddsidemargin .26in \evensidemargin .26in
        \topmargin .27in        \textwidth 145mm
        \parindent 12mm \textheight 225mm
        \headheight 0pt \headsep 0pt
        \footskip 0pt   \footheight 0pt}

%       This causes equations to be numbered by section

\def\numberbysection{\@addtoreset{equation}{section}
        \def\theequation{\thesection.\arabic{equation}}}

\def\underline#1{\relax\ifmmode\@@underline#1\else
        $\@@underline{\hbox{#1}}$\relax\fi}

%% FOLLOWING LINE CANNOT BE BROKEN BEFORE 70 CHAR
%% FOLLOWING LINE CANNOT BE BROKEN BEFORE 70 CHAR
\def\titlepage{\@restonecolfalse\if@twocolumn\@restonecoltrue\onecolumn
     \else \newpage \fi \thispagestyle{empty}\c@page\z@
        \def\thefootnote{\fnsymbol{footnote}} }

\def\endtitlepage{\if@restonecol\twocolumn \else \newpage \fi
        \def\thefootnote{\arabic{footnote}}
        \setcounter{footnote}{0}}  %\c@footnote\z@ }

\catcode`@=12
\relax

\begin{titlepage}
\hfill
\vbox{
    \halign{#\hfil         \cr
           CERN-TH/2002-345 \cr
           TAUP-2718-02 \cr
           hep-th/0211261  \cr
           } %  end of \halign
      }  % end of \vbox
\vspace*{20mm}

\begin{center}
{\Large {\bf Perturbative Computation of Glueball Superpotentials
for  SO(N) and USp(N)}\\} 
\vspace*{15mm}
%\vspace*{1mm}
{\sc Harald Ita}$^{a}$
\footnote{e-mail: {\tt ita@post.tau.ac.il}} 
{\sc Harald Nieder}$^{a}$
\footnote{e-mail: {\tt harald@post.tau.ac.il}} 
and {\sc Yaron Oz}$^{a\,b}$ 
\footnote{e-mail: {\tt yaronoz@post.tau.ac.il, Yaron.Oz@cern.ch}}\\
\vspace*{1cm} 
{\it {$^{a}$ Raymond and Beverly Sackler Faculty of Exact Sciences\\
School of Physics and Astronomy\\
Tel-Aviv University , Ramat-Aviv 69978, Israel}}\\ 

\vspace*{5mm}
{\it {$^{b}$Theory Division, CERN \\
CH-1211 Geneva  23, Switzerland}}\\

\end{center}

\vspace*{8mm}

\begin{abstract}
We use the  superspace method of hep-th/0211017 to prove
the matrix model conjecture for  {\cal N}=1 USp(N) 
and SO(N) gauge theories in four dimensions.
We derive the prescription to relate the matrix model
to the field theory computations.
We perform an explicit calculation
of glueball superpotentials. The result is consistent
with
field theory expectations.

\end{abstract}
\vskip 1cm

November 2002

\end{titlepage}

\setcounter{footnote}{0}

\phantom{\cite{Dorey:2002tj,Dorey:2002jc,Ferrari:2002jp,Fuji:2002wd,Berenstein:2002sn}}
\def\baselinestretch{1.2}
\baselineskip 16 pt
\noindent
\section{Introduction}
In a recent series of seminal papers  \cite{Dijkgraaf:2002fc,Dijkgraaf:2002vw,Dijkgraaf:2002dh}
Dijkgraaf and Vafa  proposed a  
matrix model approach for calculating holomorphic quantities in ${\cal N}=1$
supersymmetric field theories in four dimensions.
Their proposal has been tested in a variety of contexts \cite{Dorey:2002tj}-\cite{Berenstein:2002sn}
and extended to theories with flavors  \cite{Argurio:2002xv}-\cite{Demasure:2002sc}.
These techniques have been also applied to  four-dimensional ${\cal N}=2$ supersymmetric
gauge theories \cite{Naculich:2002hi}.
Other interesting features of this approach have been discussed
 in \cite{Dijkgraaf:2002pp,Gopakumar:2002wx}
and an analysis beyond the planar limit has been performed in  \cite{Klemm:2002pa}.
In   \cite{Cachazo:2002ry} the Konishi anomaly was used to elucidate the connection of the
effective superpotential to the matrix model.
 
A field theoretic proof of the conjecture of Dijkgraaf and Vafa was given in \cite{Dijkgraaf:2002xd}.
In this paper it has been shown that superspace techniques can be used to compute the perturbative part of the glueball super-potential in U(N) ${\cal N}=1$ gauge theories with one adjoint chiral superfield. The gauge sector was treated as background. 
As noted already by Dijkgraaf et al the line of argument 
should apply equally well to the gauge groups USp(N) and SO(N). 
The aim of this paper is to extend the known methods and verify the claim.

We consider one adjoint chiral superfield in an external gauge field 
with ${\cal N}=1$ supersymmetry. We wish to calculate the 
effective superpotential of the glueball superfield after integrating over the chiral superfield. The calculation has to be done by summing Feynman diagrams for the dynamical field. The Feynman diagrams can be organized in an expansion 
in terms of a small glueball superfield. We prove that as in the case of a U(N) gauge group
the computation of the Feynman graphs simplifies to graphs of a 
zero dimensional field theory, as proposed in \cite{Dijkgraaf:2002dh}. 

However, the precise prescription to relate the matrix model partition function 
to the effective super-potential has to be adapted.
Applying our results, we will calculate the perturbative superpotential up 
to order three in the glueball field for a quartic tree level superpotential.
We will find a numerical disagreement with the original conjecture.

The paper is organized in seven sections. In section two we define the theory and the observable we want to study. We then summarize briefly the superspace calculations of \cite{Dijkgraaf:2002xd}. In section three we review the convenient language for the perturbative calculations. In section four we argue that only Feynman graphs (in double line notation) on $S^2$ and $RP^2$ contribute to the perturbative calculations. In section five we prove that the field theory calculation can be reduced to a matrix model computation. The explicit prescription to relate field theory and matrix model computations is then discussed in section six. The final section contains an example computation as well as a discussion of the result. 
In the appendix we have included details about the double line notation for $SO(N)$ and $USp(N)$ and sample calculations with methods to calculate Feynman graphs with insertions developed and used in section four and five.
\section{Review of Superspace}
We start with a short review of the superspace calculation performed
in \cite{Dijkgraaf:2002xd}. This calculation does not depend on the gauge group and we will therefore
restrict ourselves to those aspects which are most important for our subsequent analysis.  

The starting point is the following 
four dimensional action
for a massive chiral superfield $\Phi$ coupled to 
an external gauge field with the self interaction determined by the 
gauge invariant super-potential $W(\Phi)$
\begin{equation}
 \label{eq:Action1}
 S(\Phi,\bar{\Phi})= \int d^4x d^4 \theta \bar{\Phi} e^V \Phi +
 \int d^4x d^2 \theta W(\Phi) +h.c. \ . 
\end{equation}
where, following the conventions given in \cite{Gates:nr}, the gauge field strength is given by
\begin{eqnarray}
  \label{eq:gau.fie.st}
  {\cal W}_{\alpha}=i\bar{D}^2e^{-V}D_{\alpha}e^{V}.
\end{eqnarray}
We are looking for the perturbative part\footnote{Here perturbative means perturbative in the field $S$, as is common in the literature on this subject.} of the effective superpotential of this system 
\begin{eqnarray}
  \label{eq:eff.super.pot.}
  \int d^2\theta\; W_{eff}^{pert.}(S)
\end{eqnarray}
as a function of the (traceless) external glueball superfield 
$S\sim{\cal W}_{\alpha}{\cal W}^{\alpha}$.  The full effective superpotential also includes the Veneziano-Yankielowicz term \cite{Veneziano:1982ah}. As we will see $W_{eff}^{pert.}$ can be calculated from summing a certain subset of Feynman graphs. 

For the perturbative analysis we should take into account
the holomorphic propagator $\left< \Phi \Phi \right>$ as well as the
anti-holomorphic propagator $\left< \bar{\Phi} \bar{\Phi} \right>$ and the mixed 
propagator $\left<\Phi \bar{\Phi} \right>$. However, as argued in \cite{Dijkgraaf:2002xd},
by holomorphy arguments we are led to the conclusion that the superpotential
for a chiral glueball superfield cannot depend on the coefficients of the
anti-chiral superpotential and thus only holomorphic propagators contribute.
 
This was put to use in \cite{Dijkgraaf:2002xd} by choosing the following 
 simple quadratic form for the
anti-chiral superpotential     
\begin{equation}  
  \label{eq:ACSP1} 
  \bar{W} \left( \bar{\Phi} \right)=\frac{1}{2} \bar{m} \bar{\Phi}^2 \ .
\end{equation}  
This choice allows to integrate out the anti-chiral superfield $\bar{\Phi}$.
Furthermore by taking into account that we are looking for the 
effective potential of a constant glueball superfield $S \sim 
{\mathcal W}^{\alpha} {\mathcal W}_{\alpha}$ it is shown in \cite{Dijkgraaf:2002xd}
that one ends up with the action  
\begin{equation}   
  \label{eq:Action2}  
  S= \int d^4x d^2 \theta \left(- \frac{1}{2 \bar{m}} \Phi
\left(\square -i {\mathcal W}^{\alpha} D_{\alpha} \right)\Phi 
+W_{tree}\left(\Phi \right)\right),
\end{equation}  
where $\bar{m}$ is the mass term that appeared in the anti-chiral superpotential (\ref{eq:ACSP1}).
The chiral tree-level superpotential will take the form 
\begin{equation} 
  \label{eq:CSP1} 
  W_{ tree}\left(\Phi \right) = \frac{m}{2} \Phi^2 + interactions \ .
\end{equation} 
As was mentioned above and as is shown explicitly in \cite{Dijkgraaf:2002xd}
the path integral does not depend on $\bar{m}$ and we will set $\bar{m}=1$.
This turns out to be the most convenient value for $\bar{m}$ if we want to
follow the approach of \cite{Dijkgraaf:2002xd} and evaluate the 
partition function perturbatively. This leads to 
\begin{equation} 
  \label{eq:Propa1}
  \frac{1}{p^2+m+{\mathcal W}^{\alpha} \pi_{\alpha}}
\end{equation}  
for the momentum space propagator. Note that after Fourier transforming the fermionic
superspace directions the derivative $D_{\alpha}$ is given by the fermionic momentum $\pi_{\alpha}$
\begin{equation}  
  \label{eq:FM1}
  D_{\alpha}=-i \pi_{\alpha}. \ .
\end{equation}  
\section{The Perturbative Computation} 
  As was shown for planar diagrams in \cite{Dijkgraaf:2002xd}, Feynman amplitudes  for the observable (\ref{eq:eff.super.pot.}) in the field theory (\ref{eq:Action1}) with gauge group $U(N)$ and $\Phi$ in the adjoint representation simplify drastically, when written in terms of Schwinger parameters. Actually, the simplification allows to reduce calculations to that of matrix models. Similar reasoning will be possible for the gauge groups $SO(N)$ and $USp(N)$. In this section we will introduce the Schwinger parameterization, perform integrations over bosonic loop momenta and point out the peculiarities of the fermionic momentum integration.  
 
Given a Feynman graph, we can write its propagators (\ref{eq:Propa1}) as integrals over Schwinger parameters $s_i$, where the index $i$ runs over the edges of the Feynman diagram,
\begin{eqnarray}
  \label{eq:schw.par.}
  \int_0^{\infty}ds_i\;exp\left[-s_i(p_i^2+{\cal W}^{\alpha}_{i}\pi_{i\alpha}+m)\right].
\end{eqnarray} 
Now the momentum integrals are Gaussian integrals and can be evaluated. To this end we express the momenta $p_i$, that flow through the i-th propagator in terms of loop momenta $k_a$ \footnote{Throughout this letter we will use the following conventions for indices: indices $i,j,...$ denote the various propagators, indices $a,b,...$ denote loops and the indices $m,n,...$ denote index loops.}
\begin{eqnarray} 
  \label{eq:loo.mom.}
  p_i=\sum_a L_{ia}k_a \ .
\end{eqnarray} 
With the conventions
\begin{eqnarray} 
  \label{eq:con.mat}
  M_{ab}(s)=\sum_is_iL_{ia}L_{ib}
\end{eqnarray} 
the integral over the bosonic momenta in all loops can be performed
\begin{eqnarray} 
  \label{eq:sch.act.}
 Z_{boson}=
\int
\prod_{a=1}^l\frac{d^4k_a}{(2\pi)^4}\;exp\left[-\sum_{a,b}k_aM_{ab}(s)k_b\right]=
\frac{1}{(4\pi)^{2l}}\frac{1}{(det\;M(s))^2} \ .
\end{eqnarray} 
It will turn out later, that the factor $(det\;M(s))^{-2}$ gets canceled by the integral over the fermionic momenta $\pi$.  

Also, we integrate over the fermionic loop momenta. With the relation
\begin{eqnarray}
  \label{eq:ganzegal}
  \pi_{i \alpha}=\sum_{a} L_{ia} \kappa_{a \alpha}
\end{eqnarray}
between the fermionic momentum $\pi_{i \alpha}$ running through the i-th propagator and the
fermionic loop momentum $\kappa_{a \alpha}$, 
the  fermionic integral takes the form
\begin{eqnarray}
  \label{eq:Ferm.Lab.}
  \int\prod_ad^2\kappa_a \; exp\left[-\sum_is_i\left(\sum_a{\cal W}^{\alpha}_iL_{ia}\kappa_{a\alpha}\right)\right] \ .
\end{eqnarray}
The fields ${\cal W}_{i \alpha}=\sum_A {\cal W}_A^{\alpha} T_{Rep(i)}^A$ are matrix-valued 
and are inserted on the i-th edge of the Feynman diagram in the representation $Rep(i)$
of the field propagating through this edge.
From now on we will consider one propagating field transforming in the adjoint representation. 

\section{Reduction to $S^2$ and $RP^2$ Graphs.}

In this section we will review why only the graphs on $S^2$ and $RP^2$ contribute to the effective superpotential. We will give the explicit form of the Feynman amplitudes that contribute to the superpotential for $SO(N)$ and $USp(N)$.

\subsection{Why $S^2$ and $RP^2$ Graphs.}

To understand why only $S^2$ and $RP^2$ graphs contribute we have to consider the ${\cal W}^{\alpha}$ insertions and the fermionic integrals (\ref{eq:Ferm.Lab.}) in more detail.

The 't Hooft double line notation helps to handle the insertions ${\cal W}^{\alpha}$ for the $SO(N)$ and $USp(N)$ gauge groups.
However, for these groups the propagators are represented not only by parallel lines, 
but also by crossed lines. 
As a consequence of the crossed lines 't Hooft 
diagrams can be associated to orientable 
and non orientable Riemann surfaces.

The 't Hooft diagrams at $l$ loops have at most $h=l+1$ holes 
(including the outer boundary for planar diagrams). 
The diagrams with $l$ loops that do not have less than $l$ 
holes are the planar ones and the non orientable ones, 
that can be drawn on $RP^2$. 
We will see shortly why these are the only relevant diagrams
for our calculation.  

We want to calculate the effective potential for the traceless glueball superfield $S$.
The perturbative expansion of the potential includes graphs with multiple insertions of
${\cal W}$ on the index loops. For example, inserting ${\cal W}$ $a_m$ times on the index loop
$m$ will give an overall factor
\begin{equation}
  \label{eq:Trace1}
  \prod_m Tr({\cal W}^{a_m}) \ ,
\end{equation}
for a given Feynman diagram. However, as we are interested in the chiral observable (\ref{eq:eff.super.pot.}), we choose the background field $\cal W^{\alpha}$ such that
the higher traces  $Tr({\cal W}^n)$ vanish for $n>2$.

The integration over the fermionic loop momenta brings down 
a factor of ${\cal W}^2$ for each loop. As follows from above, these have to be distributed in pairs to the index loops. This requires that the number of index loops be greater than or equal to the number of momentum loops.

The only oriented diagrams that meet this requirement are the planar ones, which
have l+1 index loops.
The only non-vanishing oriented diagrams are thus the planar ones
with two insertions of
${\cal W}$ on each but one index loop.

Considering the non-oriented diagrams 
for the groups $SO(N)$ and $USp(N)$ it is found that diagrams 
that can be drawn on $RP^2$ are singled out by the above requirement.   
Such diagrams with $l$ loops have $l$ index loops.
Consequently, the non-vanishing non-oriented diagrams 
have exactly two insertions of ${\cal W}$ on each index loop.

A convenient way to implement the requirement of having two insertions
for each index loop is to introduce auxiliary fermionic variables similar
to what was done in \cite{Dijkgraaf:2002xd}. However, we will assign our
auxiliary variables to the index loops and not to the loops.
A detailed description of how the appropriate transformation matrices
are obtained is given in appendix B.

Hence, we expand the background field ${\cal W}^{\alpha}_i$ as
\begin{equation}
  \label{eq:Expansion1}
  {\cal W}^{\alpha}_i=K_{im} {\cal W}^{\alpha}_m \ .
\end{equation}
Collecting the contributions of the ${\cal W}^{\alpha}$ insertions for a diagram $\gamma$ one finds the general formula
\begin{eqnarray}
  \label{eq:hilint}
  Z^{\gamma}_{ferm.}&\!\!\!\!:=\!\!\!\!&\int\prod_{a,m}d^2\kappa_ad^2{\cal W}_m\;exp\left[-\sum_is_i\left(\sum_{m,a}{\cal W}^{\alpha}_mK^T_{mi}L_{ia}\kappa_{a\alpha}\right)\right]\\
  &=&det(N(s))^2,
\end{eqnarray}
where we write the s-dependence of the above integrands in a concise way as
\begin{eqnarray}
  \label{eq:mod.vol.}
  N(s)_{ma}=\sum_is_iK^T_{mi}L_{ia} \ .
\end{eqnarray}
The s-dependence of the Feynman amplitude is then given by
\begin{eqnarray}
  \label{eq:sdepp}
  Z=Z_{boson}Z_{ferm.}=\frac{1}{(4\pi)^{2l}}\left(\frac{det(N(s))}{det(M(s))}\right)^2
  \ .
\end{eqnarray}

\subsection{Perturbative Superpotential}

Before we turn to the evaluation of $Z$ we pause for a while to write the
Feynman amplitudes in terms of the notation we have developed so far.

We use the definition for the glueball superfield 
\begin{eqnarray}
  \label{eq:gluesup}
  S=\frac{1}{32\pi^2}Tr[{\cal W}^{\alpha}{\cal W}_{\alpha}] \ ,
\end{eqnarray}
to express the amplitudes in terms of $S$. The amplitude corresponding
to a planar diagram $\gamma$ can then be written as 
\begin{eqnarray}
  \label{eq:gra.cond}
  A_{planar}^{\gamma}=N\;h\;c_{\gamma}\left(16 \pi^2  S\right)^{h-1} \int \prod_i ds_i\, e^{-s_i m}
Z(\gamma,s) \ .
\end{eqnarray}
$N$ comes from the summation over the free index loop. This factor is the same as for $U(N)$. The factor $h$ is combinatorial and counts the number of ways to pick one free index loop out of a total of $h$. Finally, $c_{\gamma}$ denotes the multiplicity of the graph $\gamma$.

For a diagram $\gamma$ on $RP^2$ we have
\begin{eqnarray}
  \label{eq:gra.con1}
  B^{\gamma}_{RP^2}=\sigma\;c_{\gamma}\left(16 \pi^2S\right)^{h}
\int \prod_i ds_i\, e^{-s_i m}
Z(\gamma,s) \ .
\end{eqnarray}
This general amplitude exhibits some important differences compared to the planar amplitude 
(\ref{eq:gra.cond}). First, we do not have any index loop without insertions.
Therefore, the factor $N$, that we have in the planar case is absent. 
Secondly, the factor $\sigma$, which takes the values $\pm1$, serves to distinguish the group 
$SO(N)$ from $USp(N)$. 
This is due to the fact that the propagator for the adjoint of $USp(N)$
also includes the skew-symmetric $USp(N)$ invariant $J$ whose
insertions give an overall minus sign for the $USp(N)$ as compared to $SO(N)$ for the $RP^2$ diagrams. The details are given in appendix A.

In terms of these amplitudes the perturbative part of the effective
superpotential is given as a function of $S$ as
\begin{eqnarray}
  \label{eq:pert.part1.}
  W^{pert.}_{eff}(S)=\sum_{\gamma}( A^{\gamma}_{planar}+B^{\gamma}_{RP^2}) \ . 
\end{eqnarray}
The amplitudes can be generated from a zero-dimensional field theory if 
$Z(\gamma,s)$ is independent of $s$.
This has been shown for $U(N)$ gauge group in \cite{Dijkgraaf:2002xd}. Similarly the s-dependence vanishes for $SO(N)$ and $USp(N)$, as we will show in the next section.

\section{Proof Of Matrix Model Conjecture}

In this section we calculate the value of the ratio $\lambda$ of the fermionic to the bosonic determinant,
\begin{eqnarray}
  \label{eq:fermbosrati}
  \lambda=\left(\frac{det(N(s))}{det(M(s))}\right)^2.
\end{eqnarray}
We will show that the ratio $\lambda$ does not depend on the Schwinger parameters $s_i$ and we will calculate its value explicitly for an arbitrary Feynman graph on $S^2$ as well as on $RP^2$. We find that $\lambda=1$ on $S^2$ and $\lambda=4$ on $RP^2$. 

For the following it will be convenient to think of $RP^2$ as a disk with opposite points on the boundary identified.

In order to actually calculate (\ref{eq:fermbosrati}) we will now relate 
the matrix $L$ to the matrix $K$, by 
introducing index loop momenta $b_m$ through $\vec{k}={\cal O}\vec{b}$. 
We find the simple relation $K=L{\cal O}$, which allows to calculate $\lambda$, 
\begin{eqnarray}
  \label{eq:fermbosrati2}
  \lambda=det({\cal O})^2.
\end{eqnarray}
That is, the Schwinger parameters cancel.

Furthermore, we will show that $det({\cal O})=2$  holds for all graphs on $RP^2$, whereas $det({\cal O})=1$ for all graphs on $S^2$ .

\begin{figure}[ht]
\begin{center}       
 \includegraphics[width=3in]{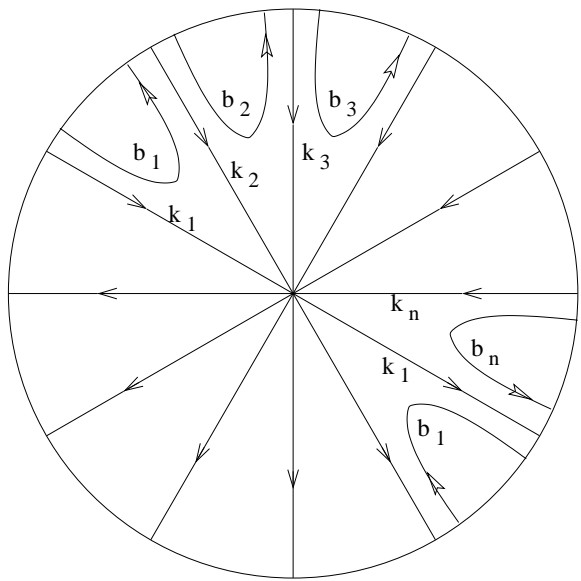}
\end{center}
\caption{{\it Star-diagram for generic graph on $RP^2$. The labels $b_i$ denote the momenta associated to the index loops. The loop momenta are denoted by $k_a$.}}
\label{stardiag}
\end{figure}

The problem of calculating ${\cal O}$ factorizes into two sub-problems: An arbitrary graph on $RP^2$ looks like a planar graph with lines emerging from the boundary loop. These lines cross the boundary of the disk, enter the disk at the opposite boundary point and, finally, fuse the boundary loop again. So a graph actually separates into two parts: The boundary loop with the emerging lines and the interior of the planar graph. Suppressing the interior, the generic exterior part is shown in Figure \ref{stardiag}.
The two parts of the graphs can be treated separately. Of course, graphs on $S^2$ are treated implicitly, by suppressing the contribution of the exterior graph.

\subsection{The Interior Graph.}
As just mentioned the interior graph looks exactly like a planar
graph, 
with the outer loop omitted. Now in planar graphs the interior index momenta $\vec{b}$ are exactly the loop momenta $\vec{k}$ such that
from
\begin{eqnarray}
  \label{eq:planloopind}
  p_i=L_{ia}k_a=K_{im}b_m \ ,
\end{eqnarray}
we see that $K=L$.
The index $i$ runs over propagators that do not cross the boundary of the disk, and the indices $a$ and $m$ label the loops and the index lines, which are actually the same.

We use the matrix ${\cal O}$ introduced above for 
the basis change from the loop momenta $k_a$ to the index momenta $b_m$,
\begin{eqnarray}
  \label{eq:basechangeo}
  b_m={\cal O}^{-1}_{ma}k_a \ .
\end{eqnarray}
According to the factorization into interior and exterior graph ${\cal O}$ splits as
\begin{eqnarray}
  \label{eq:osplitt}
  {\cal O}=\left(\begin{array}{cc}{\cal O}_{int.}&0\\0&{\cal
  O}_{ext.}\end{array}\right) \ .
\end{eqnarray}
For the interior momenta we just argued that the matrix ${\cal O}_{int.}$ obeys ${\cal O}_{int.}={\mbox{\boldmath $1$}}$. 
Now we turn to the calculation of  ${\cal O}_{ext.}$.

\subsection{The Exterior Graph}  
By definition (\ref{eq:Expansion1}) the index loop momenta $b_m$ are related to the propagator momenta $p_i$ like
\begin{eqnarray}
  \label{eq:propindloop}
  p_i=K_{im}b_m \ .
\end{eqnarray}
Here $m,i$ run from $1$ to $n$, the number of twisted propagators.
\footnote{Twisted propagators are propagators that cross the boundary of the disk.}
The index loop momenta, on the other hand, give the loop momenta $k_a$. As can be read off from Figure \ref{stardiag}, the momenta flowing in the index lines $b_i$ are related to the momenta in the canonical loops $k_i$ like 
\begin{eqnarray}
  \label{eq:momentumtrafo}
  k_1&=&b_1+b_n,\\
  k_l&=&b_l-b_{l-1},\quad l\neq 1 \ .
\end{eqnarray}
Thus we find the identity 
\begin{eqnarray}
  \label{eq:lko}
  K{\cal O}^{-1}=L \ .
\end{eqnarray}
Plugging this relation into the definitions of $M(s)=L^TSL$ and $N(s)=K^TSL$, we find, as promised
\begin{eqnarray}
  \label{eq:lameq}
  \lambda=det({\cal O})^2 \ .
\end{eqnarray}
It is easy to calculate the determinant of ${\cal O}$ for arbitrary $n$. The value is 
\begin{eqnarray}
  \label{eq:trans.mat.}
  det({\cal O})=1+(-1)^{n+1}(-1)^{n-1}=2 \ .
\end{eqnarray}
Hence, the value of the factor $\lambda$ is
\begin{eqnarray}
  \lambda=\left(\frac{det(N(s))}{det(M(s))}\right)^2=det({\cal O})^2=4
  \ .
\end{eqnarray}

\section{Discussion of Matrix Model Conjectures}

After the cancellation of the fermionic and bosonic measures,
 the integrals over the Schwinger parameters (\ref{eq:Propa1}) are given by
\begin{eqnarray}
  \label{eq:feynampsch}
  \int\prod_{i=1}^Lds_i e^{-s_im}=\frac{1}{m^L} \ .
\end{eqnarray}
(L denotes the number of propagators, i.e. links in a given Feynman graph.)
This factor can be reproduced by a mass term in the zero dimensional action $\frac{m}{2}Tr[\Phi^2]$. The combinatorial factors are reproduced by the Feynman rules of a zero dimensional field theory, i.e. a matrix model. 

It was conjectured in \cite{Dijkgraaf:2002dh} and later shown in
\cite{Dijkgraaf:2002xd,Cachazo:2002ry} that for the gauge group $U(N)$ the 
perturbative part of the effective superpotential
$W_{eff}(S)$ can be obtained from a matrix model calculation as
\begin{eqnarray}
  \label{eq:pert.Suppo.un}
  W_{eff}^{pert.}(S)=N\frac{\partial {\cal F}_0(S)}{\partial S} \ ,
\end{eqnarray}
where ${\cal F}_0$ is the planar contribution to the free energy of the $U(N)$ matrix model with potential
equal to the tree level potential of the gauge theory. The $S$ dependence
of ${\cal F}_0$ is induced by identifying the 't Hooft coupling
$gN$ of the matrix model with the glueball-field $S$.

It was anticipated in \cite{Dijkgraaf:2002dh} that for the gauge groups
$SO(N)$ and $USp(N)$ one also has to take into account the contribution
of the non-oriented matrix model diagrams. The above analysis suggests the form
\begin{eqnarray}
  \label{eq:pert.Suppo.}
  W_{eff}^{pert.}(S)=N\frac{\partial {\cal F}_0(S)}{\partial S}+\lambda{\cal G}_0(S) \ ,
\end{eqnarray}
which differs from the conjecture \cite{Dijkgraaf:2002dh} of the non-perturbative superpotential, as a factor $N$ not $(N\mp2)$ multiplies the derivative of ${\cal F}_0(S)$. Also we find a multiplicative factor in front of ${\cal G}_0$, the $RP^2$ part of the matrix model free energy. The factor comes from the volume of the Schwinger moduli space (\ref{eq:fermbosrati},\ref{eq:fermbosrati2}) and is given by $\lambda=4$. \footnote{As already mentioned, the full effective superpotential also includes the Veneziano-Yankielowicz term $-(N\mp2)[Slog(S/\Lambda^3)-S]$ \cite{Veneziano:1982ah}, which in the matrix model is related to the volume of the gauge group \cite{Ooguri:2002gx}.}

Our explicit calculation of the next section indicates, that for a quartic tree level superpotential, (\ref{eq:pert.Suppo.}) can be put into the form of (\ref{eq:pert.Suppo.un}) with the non-oriented part accounting for a shift $N \to N \mp 2$. We will comment on this case below.
   
\section{Perturbative Superpotential Computations}
In this section we explicitly calculate the effective superpotential for the gauge groups $SO(N)$ and $USp(N)$ with quartic tree level potential. We compare the contribution of non-orientable Feynman diagrams and planar ones. 
We will find a relation between the planar and the non-oriented contributions leading to a simplification
of the matrix model description (\ref{eq:pert.Suppo.}).

\subsection{General Prescription}

With the results of the preceding sections at hand we can now rewrite the 
amplitudes in (\ref{eq:pert.part1.}).
We get
\begin{eqnarray}
  \label{eq:gra.cond.zus}
  A_{planar}^{\gamma}=N\;h\;c_{\gamma} \left( \frac{1}{m} \right)^L S^{h-1}
\end{eqnarray}
for a given planar Feynman diagram $\gamma$. 
For the non-oriented diagrams we obtain
\begin{eqnarray}
  \label{eq:gra.con1.zus}
  B^{\gamma}_{RP^2}=\sigma\lambda_{RP^2}\;c_{\gamma}\left( \frac{1}{m} \right)^L S^{h} \ .
\end{eqnarray}
We note that the powers of $16 \pi^2$ in (\ref{eq:gra.cond},\ref{eq:gra.con1}),
are canceled by $(4 \pi)^{2l}$ coming from the bosonic momentum integration (\ref{eq:sch.act.}).
(Remember that the number of loops $l$ is related to the number of index loops $h$ by $l=h-1$ 
for the planar and $l=h$ for the non-oriented diagrams.)
\subsection{Explicit Calculations}

\begin{figure}[h]
\begin{center}
 \includegraphics[width=4in]{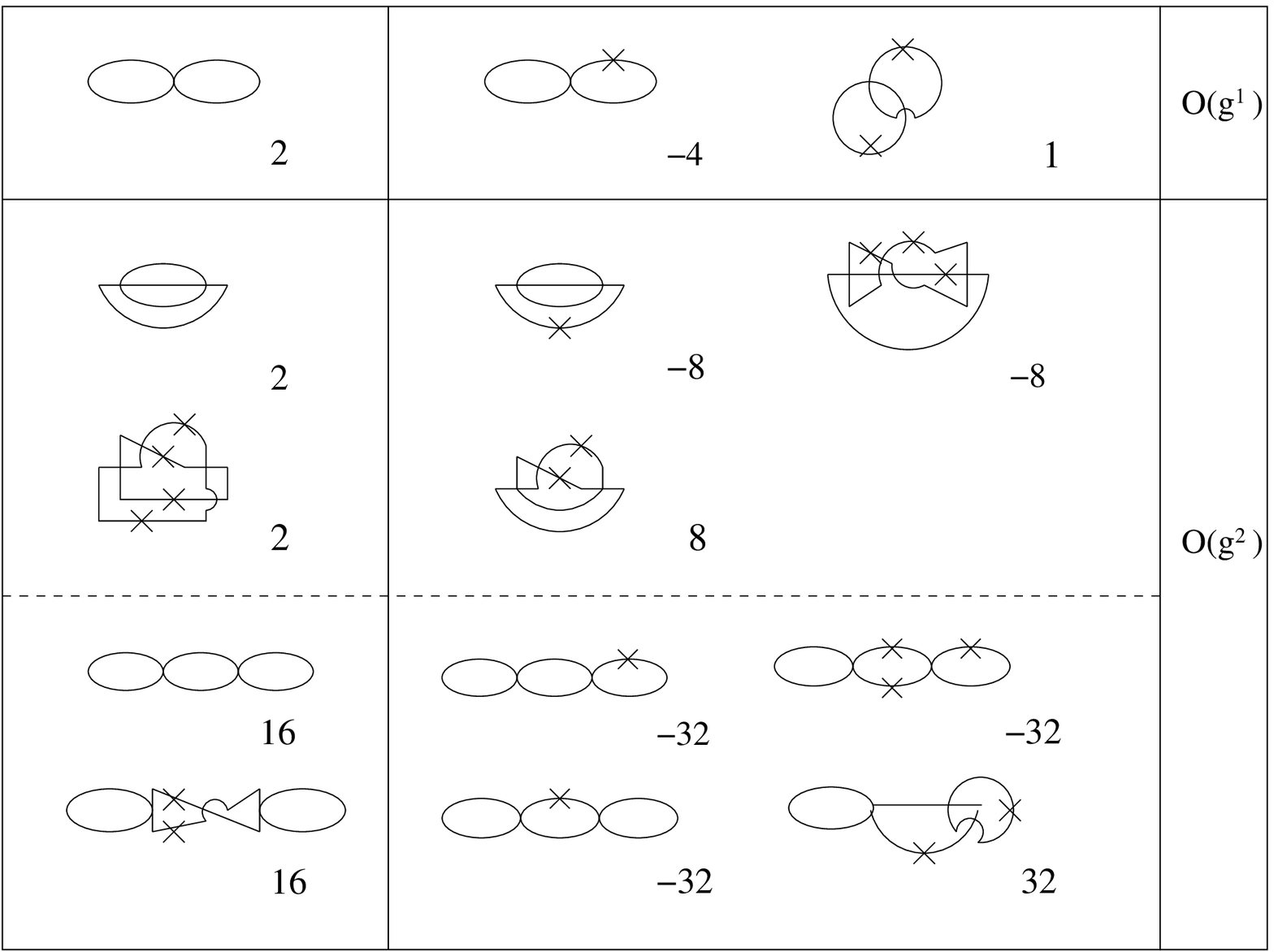}
\end{center}
\caption{{\it Feynman graphs for perturbative calculation in $W_{tree}(\Phi)=\frac{m}{2}Tr \Phi^2+2g\; Tr\Phi^4$ up to two vertices.}}
\label{gr.or.}
\end{figure}
We explicitly apply the rules from above to calculate 
the perturbative part of the super-potential for the tree level  super-potential 
\begin{eqnarray}
  \label{eq:suppot}
  W_{tree}(\Phi)=\frac{m}{2}Tr \Phi^2+2g\; Tr\Phi^4 \ .
\end{eqnarray}
The combinatorial factors $c_{\gamma}$ in equations 
(\ref{eq:gra.cond.zus},\ref{eq:gra.con1.zus}) receive contributions  \footnote{L,V and F will in the following denote the number of propagators, i.e. links, vertices and faces of a given Feynman graph. Furthermore we will use the relation $2V=L$ for the $\Phi^4$ interaction.}  
$(1/2)^L$ from the propagators (\ref{eq:propfree},\ref{eq:propfree2})
and a factor of $2g$ from each vertex. The total is given by
\begin{eqnarray}
  \label{eq:twistfac}
  c_{\gamma}&=&\left(\frac{1}{2}\right)^L (2g)^V\times (combinatorics)
  \ .
\end{eqnarray}
The superpotential is then given by
\begin{eqnarray}
  \label{eq:superpotlead}
  W_{eff}^{pert.}(S)&=&N[3S^2 \left( \frac{g}{m^2} \right)
+36S^3\left( \frac{g}{m^2} \right)^2+...]  \\ \nonumber 
 & & +\sigma\lambda_{RP^2}[\frac{3}{2}S^2\left( \frac{g}{m^2} \right)+\frac{36}{2}S^3\left( \frac{g}{m^2} \right)^2+...] \ ,
\end{eqnarray}
where $\sigma= \mp 1$ for the groups $SO(N)$ and $USp(N)$, respectively, and
$\lambda_{RP^2}=4$.

% we find that $W_{eff}^{pert.}(S)$ is exactly of the form
%(\ref{eq:pert.Suppo.}) with $N$ shifted to $N \mp 2$ due to the non-oriented
%diagrams, where ${\cal F}_0$ is now the free energy of a $SO(N)/USp(N)$
%matrix model with potential (\ref{eq:suppot}).
%Besides, by rescaling the coupling constant as $g \to g/8$  we see that this result
%agrees with the perturbative part of the superpotential
%obtained for the $SO(N)$, $USp(N)$ theories from geometric engineering
%in \cite{Fuji:2002vv}. 

\subsection{Discussion of Example}

The above result for $W_{eff}^{pert.}(S)$ is exactly of the form
(\ref{eq:pert.Suppo.un}) with $N$ shifted to $N \mp 2$ due to the non-oriented
diagrams, where ${\cal F}_0$ is now the planar free energy of a $SO(N)/USp(N)$
matrix model with potential (\ref{eq:suppot}). Thus, up to order
$O(S^4)$ we get 
\begin{eqnarray}
  \label{eq:pert.Suppo1.}
  N\frac{\partial {\cal F}_0(S)}{\partial S}+\lambda{\cal G}_0(S)=
(N\mp2)\frac{\partial {\cal F}_0(S)}{\partial S} \ .
\end{eqnarray}
A factorization argument presented in \cite{Fuji:2002vv} suggests that
the relation (\ref{eq:pert.Suppo1.})
is exact. 
Consider $N=2$ super Yang Mills, softly broken
to 
$N=1$ gauge theory by a quartic tree level potential. 
It was argued in \cite{Fuji:2002vv} that the value at the minimum 
of the effective superpotentials for 
$SO(2KN-2K+2)$ (or $USp(2KN+2K-2)$) gauge groups is related to the
value at the minimum of the effective superpotential of $SO(2N)$ (or $USp(2N)$) as 
\begin{eqnarray}
  \label{eq:effsuperpotjap}
  W_{eff}^{2KN\mp2K\pm2}(g,m)=KW_{eff}^{2N}(g,m) \ ,
\end{eqnarray}
for a fixed quartic potential $W_{tree}(\Phi)=m/2\,Tr \Phi^2+2g\; Tr\Phi^4$.
The value of the effective superpotential, calculated from the $N=1$
approach 
and the $N=2$ approach are related as shown in \cite{Cachazo:2002pr}.

Applying the factorization (\ref{eq:effsuperpotjap}) suggest for this example that
\begin{eqnarray}
  \label{eq:pert.Suppo.fr}
W_{eff}^{pert.}(S) = (N\mp2)\frac{\partial {\cal F}_0(S)}{\partial S} \ .
\end{eqnarray}

%{\bf topological string to be added or forgotten}
%However, the relation between ${\cal F_0}'$ and ${\cal G_0}$ is not expected to hold for other tree level potentials. It would be interesting to investigate how this comes about in the topological closed string, dual to the matrix model. Naively it looks like an non-local effect between different cohomology cycles in the deformed conifold.

\vskip 1cm

\flushleft{{\bf Acknowledgments}
We would like to thank O. Aharony, U. Fuchs, T. Sakai, J. Sonnenschein
for valuable discussions.
The research was supported by the US-Israel Binational Science Foundation.
The research of H.I.is supported by the TMR European Research Network.

\newpage
\appendix
\section{Double line notation}
\subsection{Double line notation for $SO(N)$}
It is convenient to represent the Lie algebra of $SO(N)$ by antisymmetric $N\times N$ matrices $M_{mn}$, with the condition $M_{mn}=-M_{nm}$. In this notation adjoint fields are real antisymmetric $N\times N$ matrices 
\begin{eqnarray}
  \label{eq:adjfield}
  \Phi_{mn}=-\Phi_{mn},
\end{eqnarray}
and their free propagator in momentum 
space is proportional to the projector $P_{kl}\;_{mn}=\frac{1}{2}(\delta_{km}\delta_{ln}-\delta_{lm}\delta_{kn})$,
\begin{eqnarray}
  \label{eq:propfree}
  \langle\Phi_{kl}\Phi_{mn}\rangle\sim\frac{1}{2}(\delta_{km}
\delta_{ln}-\delta_{lm}\delta_{kn}) \ .
\end{eqnarray}
In double line notation we get figure \ref{gr.itor.}.
\begin{figure}[h]
\begin{center}
 \includegraphics[width=3in]{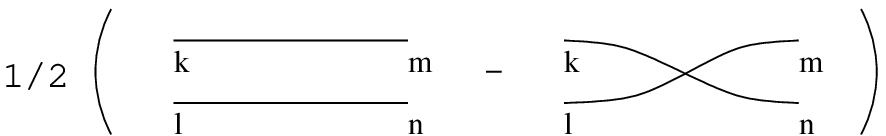}
\end{center}
\caption{{\it The propagator of SO(N) in double line notation.}}
\label{gr.itor.}
\end{figure}
The ends of the lines represent indices and the lines represent 'deltas'.

Next we will consider insertions of the gauge fields 
\begin{eqnarray}
  \label{eq:gauge insertions}
  exp(-s[{\cal W}^{\alpha},\;\cdot\;]\pi_{\alpha}) \ ,
\end{eqnarray}
in the double line notation. A simple commutator $[{\cal W},\;\cdot\;]$ gives
\begin{eqnarray}
  \label{eq:commcontr}
  \delta_{km}{\cal W}_{ln}-{\cal W}_{mk}\delta_{ln} \ .
\end{eqnarray}
Pictorially the index contractions are drawn as in figures \ref{gr.or.1}.
\begin{figure}[h]
\begin{center}
 \includegraphics[width=3in]{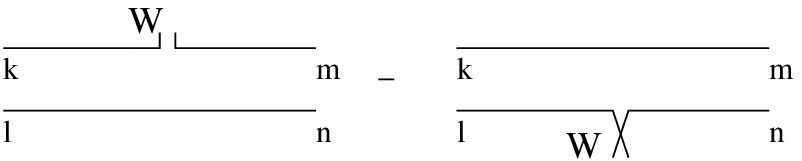}
\end{center}
\caption{{\it The insertion of a commutator  $[{\cal W},\;\cdot\;]$ of ${\cal W}\in so(N)$ in double line notation.}}
\label{gr.or.1}
\end{figure}
If we use the property ${\cal W}_{km}=-{\cal W}_{mk}$ of the 
representation (\ref{eq:adjfield}) $so(N)$, we can rewrite the commutator as
\begin{eqnarray}
  \label{eq:commcontr22}
  \delta_{km}{\cal W}_{ln}+{\cal W}_{km}\delta_{ln}
\end{eqnarray}
and in double line notation the second term is denoted by figure \ref{gr.or.2}
\begin{figure}[h]
\begin{center}
 \includegraphics[width=3in]{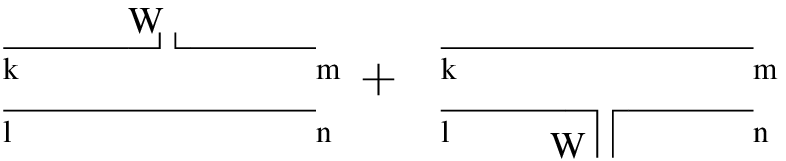}
\end{center}
\caption{{\it The insertion of a commutator  $[{\cal W},\;\cdot\;]$ of ${\cal W}\in
  so(N)$  
in double line notation, after applying the property ${\cal W}_{km}=-{\cal W}_{mk}$.}}
\label{gr.or.2}
\end{figure}
For the exponent (\ref{eq:gauge insertions}) this generalizes to
\begin{eqnarray}
  \label{eq:expinsert1}
  \delta_{km}\; \big[exp(-s{\cal W}\pi)\big]_{ln}+\big[exp(-s{\cal W}\pi)\big]_{km}\;\delta_{ln}.
\end{eqnarray}
The generalization of the above diagrams is straightforward.
\subsection{Double line notation for $USp(N)$}
It is convenient to represent the Lie algebra of $USp(N)$, $N$ even, by $N\times N$ matrices $M_{mn}$, with the condition $M_{mn}=(JMJ)_{nm}$, where J denotes the USp(N) invariant skew symmetric form
\begin{eqnarray}
  \label{eq:J}
  J=\left(\begin{array}{cc}0&{\mbox{\boldmath $1$}}_{N/2\times N/2}\\ -{\mbox{\boldmath $1$}}_{N/2\times N/2}&0\end{array}\right), J^2=-{\mbox{\boldmath $1$}}_{N\times N}.
\end{eqnarray}

The propagator of a free adjoint scalar $\Phi_{mn}$ is proportional to the projector $P_{kl}\;_{mn}=\frac{1}{2}(\delta_{km}\delta_{ln}-J_{lo}\delta_{om}J_{kp}\delta_{pn})$ 
\begin{eqnarray}
  \label{eq:propfree2}
  \langle\Phi_{kl}\Phi_{mn}\rangle\sim\frac{1}{2}(\delta_{km}\delta_{ln}-J_{lm}J_{kn}).
\end{eqnarray}
In double line notation $J$ insertion are denoted by arrows, see figure \ref{gr.or.sp}. The orientation defined by the arrows is important, since $J$ is antisymmetric and the order of its index contraction gives crucial $(-1)$ factors.
\begin{figure}[h]
\begin{center}
 \includegraphics[width=3in]{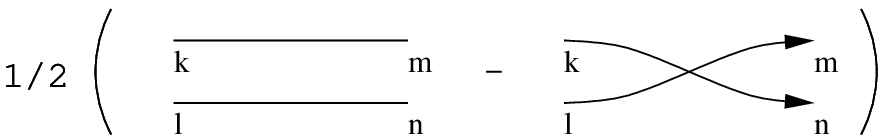}
\end{center}
\caption{{\it The propagator of an adjoint field of $USp(N)$ in double line notation.}}
\label{gr.or.sp}
\end{figure}
Insertions of a commutator can be rewritten using the properties $M_{mn}=(JMJ)_{nm}$ of $usp(N)$ to give
 \begin{eqnarray}
  \label{eq:expinsert2}
  \delta_{km}\; {\cal W}_{ln}-(J{\cal W}J)_{km}\;\delta_{ln}.
\end{eqnarray}
In double line notation these terms are shown in figure \ref{gr.ors.}.
\begin{figure}[h]
\begin{center}
 \includegraphics[width=3in]{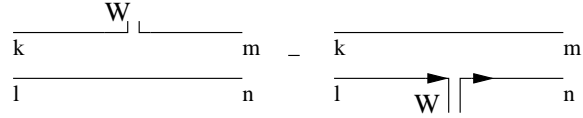}
\end{center}
\caption{{\it The insertion of a commutator  $[{\cal W},\;\cdot\;]$ of ${\cal W}\in usp(N)$  in double line notation, after applying the property $M_{mn}=(JMJ)_{nm}$.}}
\label{gr.ors.}
\end{figure}
Insertions of the gauge fields like in (\ref{eq:gauge insertions}), factors \footnote{Here we used the $exp(J{\cal W}J)=-Jexp(-{\cal W})J$} 
\begin{eqnarray}
  \label{eq:expinsert3}
  \delta_{km}\; \left[exp(-s{\cal W}\pi)\right]_{ln}+\left[Jexp(s{\cal W}\pi)J\right]_{km}\;\delta_{ln}.
\end{eqnarray}
We used the properties of the $USp(N)$ generators to exchange the indices $k\leftrightarrow m$, and got two $J$s and a minus sign. 

\section{The Matrices $L_{ia}$, $K_{im}$}
In this appendix we will use explicit examples to illustrate the algorithm to
determine the matrices $L_{ia}$ and $K_{im}$.

\subsection{Planar Diagrams}

The planar diagram we will analyze is given in figure \ref{gr1}. We denote the momenta associated to the propagators
by $p_i$ and the loop momenta by $k_a$. The ${\cal W}_i$ associated to the propagators are also written.
The empty arrows indicate the index loops, where the subscript reminds us to associate 
the auxiliary variable $\hat{{\cal W}}_m$ to the respective index loop. As explained in the text
for a planar diagram to be non-vanishing we have to choose one index loop to be free of insertions of the background field.
This choice corresponds in figure \ref{gr1} to the choice of putting the empty arrows on three out of the four index loops.

\begin{figure}[h]
\begin{center}
 \includegraphics[width=3.5in]{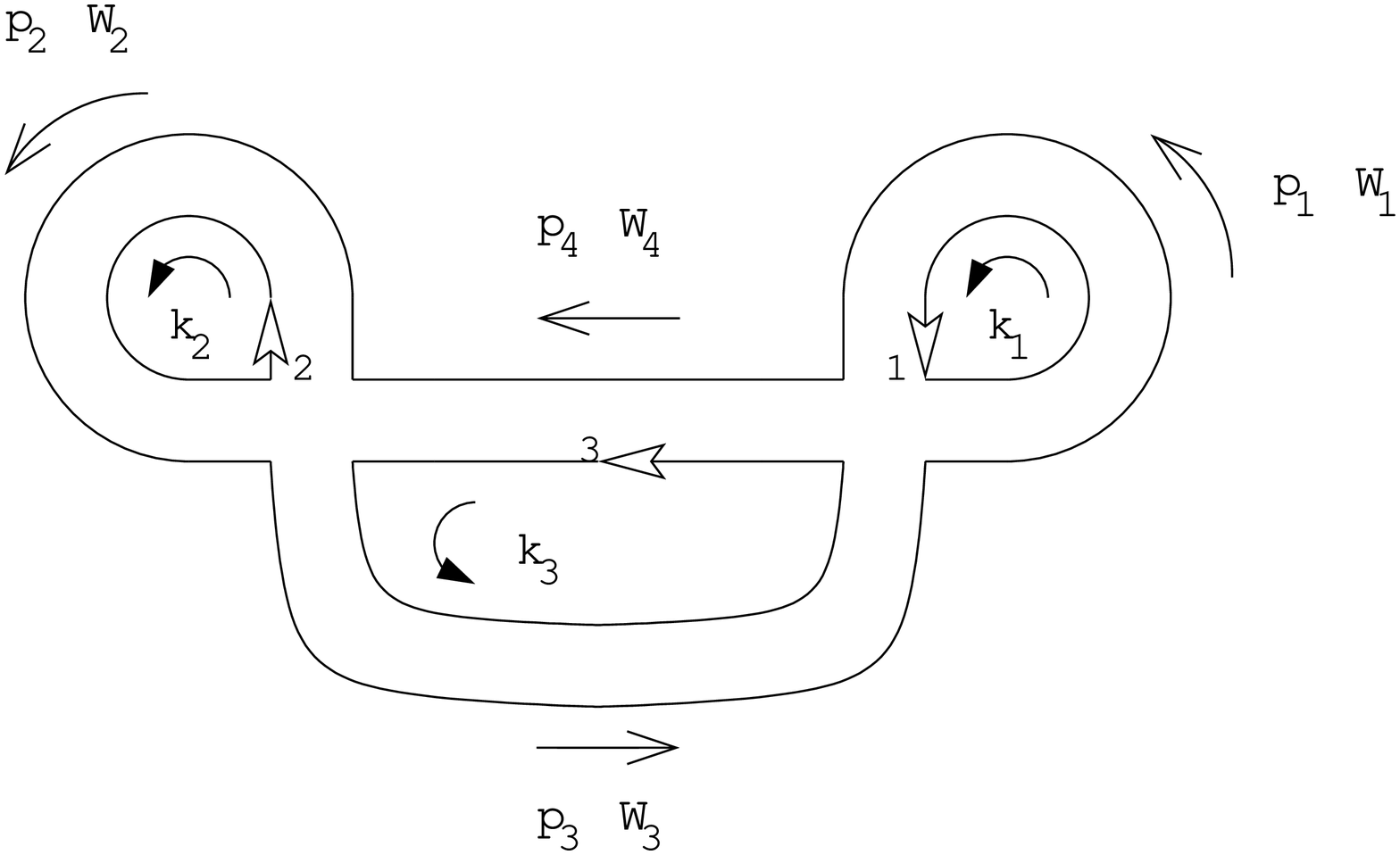}
\end{center}
\caption{{\it Planar Feynman diagram with loop momenta $k_a$, propagator momenta $p_i$, insertions $W_i$ and three out of four index loops, indicated by empty arrows. The outer index loop is chosen to be free of insertions of $W_i$.}}
\label{gr1}
\end{figure}

The matrix $L_{ia}$ gives the expansion of the propagator momenta in terms of the loop momenta. For the graph in Figure \ref{gr1}
we can make the following choice
\begin{equation}
  \label{eq:AppL1}
  p_1=k_1, \; p_2=k_2, \; p_3=k_3 \ .
\end{equation}
Furthermore momentum conservation at the vertices implies that $p_3=p_4$ and thus
$p_4=k_3$.
Hence, the matrix $L_{ia}$ is given by
\begin{eqnarray}
  \label{eq:AppL2}
  L= \left( 
    \begin{array}{c c c}
    1 & 0 & 0 \\
    0 & 1 & 0 \\
    0 & 0 & 1 \\
    0 & 0 & 1
    \end{array}
\right)
\end{eqnarray}
Some straightforward algebra then gives for the product $s_i (L^T)_{ai} L_{ib}$
\begin{eqnarray}
  \label{eq:AppLSL1}
  \left( 
    \begin{array}{c c c}
    s_1 & 0 & 0 \\
    0 & s_2 & 0 \\
    0 & 0 & s_3+s_4 
    \end{array}
\right) \ .
\end{eqnarray}
In order to obtain the matrix $K_{im}$ we just need to read off from the diagram
the contributions of each index loop to a given ${\cal W}_i$ taking into account the mutual orientation.
This is done in the following way. Starting with ${\cal W}_1$ we notice that there is only a contribution from the
first index loop. Moreover, we have chosen the orientation of the index loop to be the same as the orientation 
of $p_1$, such that we can write
\begin{equation}
  \label{eq:AppW1}
  {\cal W}_1= \hat{{\cal W}}_1 \ .
\end{equation}
A similar reasoning applies to the remaining ${\cal W}_i$ such that the matrix $K$ relating the ${\cal W}_i$
to the auxiliary variables $\hat{{\cal W}}_m$ according to ${\cal W}_i=K_{im}\hat{{\cal W}}_m$  is given by
\begin{eqnarray}
  \label{eq:AppKplanar1}
  K= \left( 
    \begin{array}{c c c}
    1 & 0 & 0 \\
    0 & 1 & 0 \\
    0 & 0 & 1 \\
    0 & 0 & 1
    \end{array}
\right) \ ,
\end{eqnarray}
which equals the matrix $L$ as claimed in  \cite{Dijkgraaf:2002xd}.

\subsection{Non-Oriented Diagrams}

\subsubsection{The first non-oriented diagram}
The first non-oriented diagram we will discuss is given in 
figure \ref{gr.nieor.}. The notation is the same as in the planar case.
The only difference is that now due to the twist in one 
of the propagators the number of index loops is the same as the number
of momentum loops and we do not have any free index loop. 

\begin{figure}[h]
\begin{center}
 \includegraphics[width=3.5in]{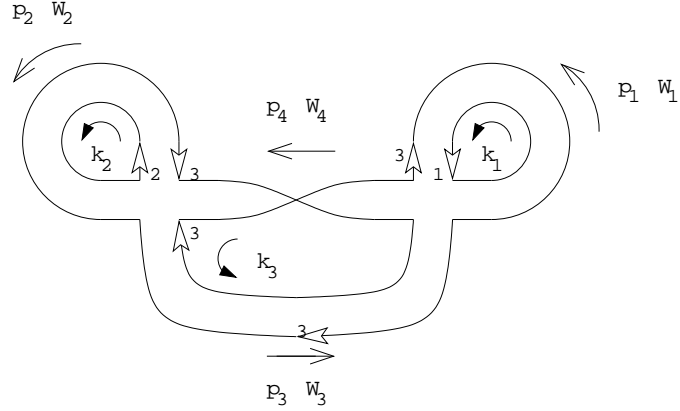}
\end{center}
\caption{{\it Non-orientable Feynman diagram with loop momenta $k_a$, propagator momenta $p_i$, insertions $W_i$ and three index loops, indicated by empty arrows.}}
\label{gr.nieor.}
\end{figure}

The matrix $L_{ia}$ is easily seen to be the same as in the planar
diagram of 
figure \ref{gr.nieor.}.
The matrix $K_{im}$ on the other hand is modified. Applying the same strategy as in the planar case we find that now 
${\cal W}_1$ will receive contributions not only from index loop 1 but also from index loop 3. The orientation
of index loop 1 is again chosen to be the same as the orientation of the momentum $p_1$, 
whereas index loop 3 is oriented in the opposite way. 
The contribution from index loop 3 will therefore come with a negative sign and 
we can write
\begin{equation}
  \label{eq:AppNOW1}
   {\cal W}_1= \hat{{\cal W}}_1 - \hat{{\cal W}}_3\ .
\end{equation}
For ${\cal W}_2$ we obtain a similar result and for ${\cal W}_3$, ${\cal W}_4$ we find that
the third index loop contributes twice at a time with a negative sign leading to
\begin{equation}
  \label{eq:AppNOW2}
  {\cal W}_{3}= -2\hat{{\cal W}}_3 = {\cal W}_4 \ .
\end{equation}
Thus, the matrix $K$ reads
\begin{eqnarray}
\label{eq:AppNOK1}
 K= \left( 
    \begin{array}{c c c}
    1 & 0 & -1 \\
    0 & 1 & -1 \\
    0 & 0 & -2 \\
    0 & 0 & -2
    \end{array}
 \right) \ ,
\end{eqnarray}
and the matrix $s_i(K^T)_{ni} L_{ib}$ is given by
\begin{eqnarray}
  \label{eq:AppNOSKL1}
  \left( 
    \begin{array}{c c c}
    s_1 & 0 & 0 \\
    0 & s_2 & 0 \\
    -s_1 & -s_2 & -2(s_3+s_4) 
    \end{array}
 \right) \ .
\end{eqnarray}
\subsubsection{The second non-oriented diagram}
Our second non-oriented diagram is drawn in figure \ref{gr.or2.}. 
\begin{figure}[h]
\begin{center}
 \includegraphics[width=3.5in]{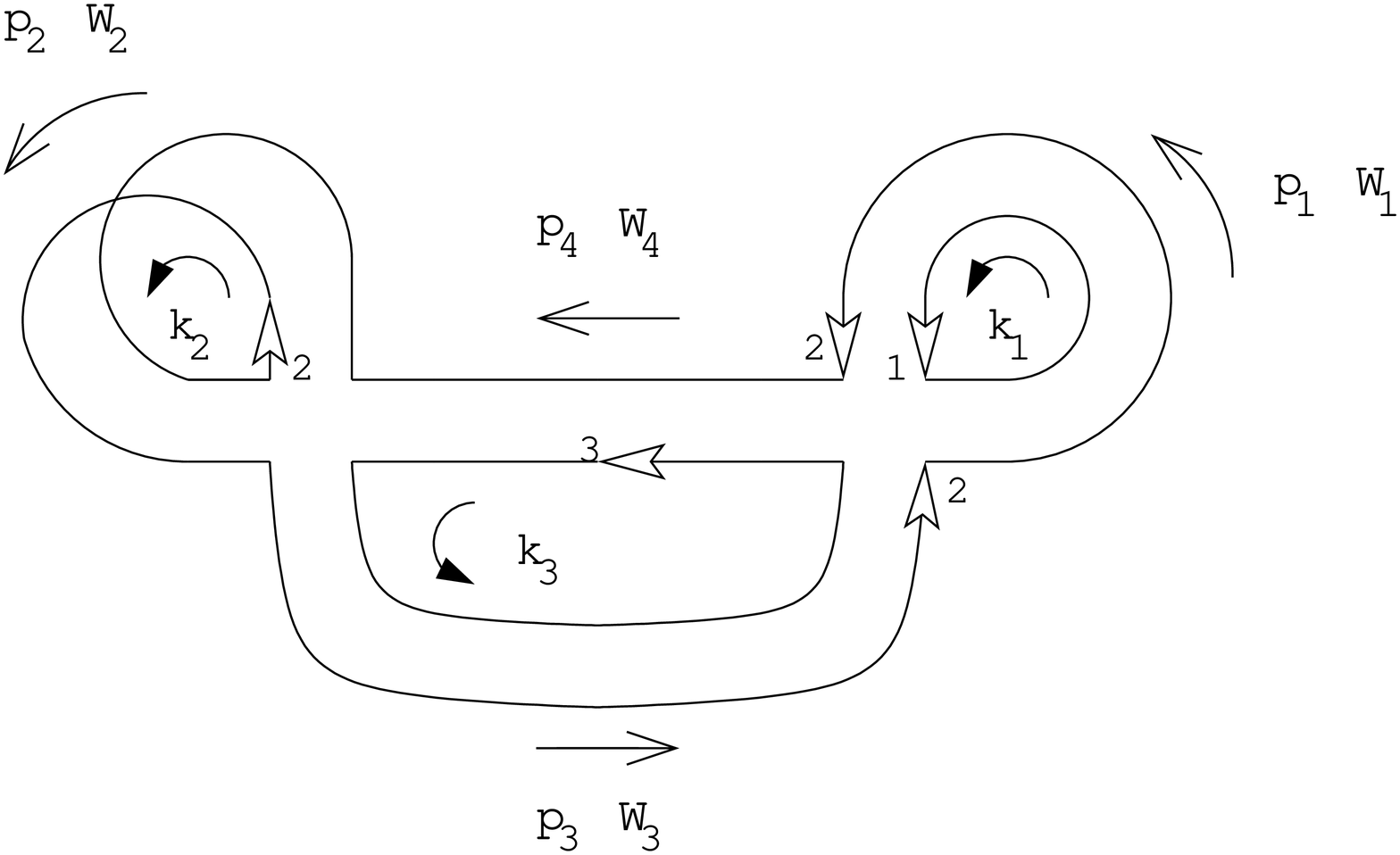}
\end{center}
\caption{{\it Non-orientable Feynman diagram with loop momenta $k_a$, propagator momenta $p_i$, insertions $W_i$ and three index loops, indicated by empty arrows.}}
\label{gr.or2.}
\end{figure}

The matrix $K$ for this diagram is again obtained by applying the above algorithm.
Reading off the contributions of the index loops we find
\begin{eqnarray}
  \label{eq:AppNOK2}
 {\cal W}_1 & = & \hat{{\cal W}}_1 + \hat{{\cal W}}_2, \\ \nonumber
 {\cal W}_2 & = &  2\hat{{\cal W}}_2 , \\    \nonumber
 {\cal W}_3 & = & \hat{{\cal W}}_2 + \hat{{\cal W}}_3, \\ \nonumber
 {\cal W}_4 & = & \hat{{\cal W}}_2 + \hat{{\cal W}}_3 \ , 
\end{eqnarray}
leading to
\begin{eqnarray}
  \label{eq:AppNOK3}
  K= \left( 
    \begin{array}{c c c}
    1 & 1 & 0 \\
    0 & 2 & 0 \\
    0 & 1 & 1 \\
    0 & 1 & 1
    \end{array}
 \right) \ .
\end{eqnarray}
In this case the matrix $s_i(K^T)_{ni} L_{ib}$ is 
\begin{eqnarray}
  \label{eq:AppNOSKL2}
  \left( 
    \begin{array}{c c c}
    s_1 & 0 & 0 \\
    s_1 & 2s_2 & s_3+s_4 \\
    0 & 0 & -(s_3+s_4) 
    \end{array}
 \right) \ .
\end{eqnarray}

\end{document}